\documentclass[12pt]{iopart}

\def\be{\beta}
\def\ga{\gamma}
\def\etall{{\it et al.}}
\def\gev{\mbox{ GeV}}
\def\cpt10{CPT~'10}
\begin{document}

\title{Testing Lorentz symmetry with atoms and light}

\author{Neil Russell}

\address{Physics Department,
Northern Michigan University,
Marquette, MI, USA}
\ead{nrussell@nmu.edu}
\begin{abstract}
This article reports on the
Fifth Meeting on CPT and Lorentz Symmetry,
\cpt10,
held at the end of June 2010 in Bloomington, Indiana, USA.
The focus is on recent tests of Lorentz symmetry
using atomic and optical physics.
\end{abstract}

\section{Introduction}
Experiments with atoms, molecules, and electromagnetic waves
have a remarkable track record of innovations
and sensitivity improvements.
As the technology has advanced,
new views of the details of nature have become possible.
Recently,
atom interferometers have probed the gravitational redshift
\cite{2011equivalencePRL},
comagnetometers have sought spin couplings at unprecedented levels
\cite{10Brown},
antihydrogen has been cornered in an electromagnetic trap
\cite{2010Andresen},
and rotating optical resonators have
sought delicate anisotropies
in the speed of light
\cite{09Eisele,09Herrmann}.

These experiments have created
new opportunities for exploring
the fundamental symmetries of nature.
One area is Lorentz symmetry,
the hypothesis that experiments give the same results
regardless of their inertial reference frame
and regardless of their orientation.
Closely related is
the hypothesis of CPT symmetry,
that the outcome of an experiment
must be the same as that of its CPT image
\cite{02greenberg}.
This image is constructed
by replacing each charge with its conjugate,
by parity-reversing the entire physical configuration,
and
by reversing all initial velocities
so that time runs `backwards.'
These two fundamental symmetries
have withstood hundreds of tests
spanning decades,
although recently
anomalous effects
with mesons and neutrinos
have provided hints of possible symmetry breaking
\cite{2010AKRvK,puma}.

The discovery of Lorentz and CPT symmetry breaking
would necessitate fundamental changes in
the conventional theories of gravity and quantum mechanics.
It would cause a stir in the community,
since the nature of the violation would
most likely hold essential information about
a realistic quantum-gravity theory.
Since violations of these symmetries
have never been seen,
they are likely to be minuscule
or countershaded by weak-gravity couplings.
One challenge of detecting them
is attaining sufficient experimental sensitivity
to signals.
Another is designing experiments
with sensitivity to likely types of unconventional signal,
and
this implies the need for a theoretical framework
that incorporates Lorentz and CPT violations.
The Standard-Model Extension (SME),
developed over a period of more than 20 years,
and first published in basic form
in the mid 1990s,
does just this,
providing a detailed theoretical framework
containing all possible Lorentz and CPT breaking
terms at the level of the fundamental particles and
interactions \cite{dcak97}.
The deep interest in exploring Lorentz and CPT symmetry
can be attributed to the confluence of these circumstances --
strong theoretical motivations,
advancing experimental techniques,
and the multitudes of measurable coefficients for Lorentz and CPT violation.
This interest was clearly evident at the
{\em Fifth Meeting on Lorentz and CPT Symmetry,}
\cpt10,
held in Bloomington, Indiana,
at the end of June 2010.

This conference report summarizes the content
of the \cpt10 meeting,
focusing in particular on AMO physics.
Since the first meeting on Lorentz and CPT Symmetry,
held in Bloomington in 1998,
the number of experimental results
measuring SME coefficients has snowballed.
The full listing spans the sectors of physics
and is growing yearly \cite{LVtables}.
Details of individual presentations can be found
in the conference proceedings,
volume V in the series \cite{proceedings}.

The SME is a realistic effective field theory
in which the coefficients for Lorentz breaking
could arise from spontaneous symmetry breaking
in a fundamental theory such as string theory
\cite{1989ssb}.
It exploits the fact that the effects of such symmetry violations
could be detected at attainable energies in the context of effective field theory
\cite{1995eft}.
In the first \cpt10 talk,
distinguished physicist James Bjorken,
of the SLAC National Accelerator Laboratory,
discussed these and other ideas surrounding spontaneously-broken symmetries.
He provided a historical perspective on this field
and discussed contributions made for example
by Nobel laureate Yoichiro Nambu,
who addressed the
2001 and 2004 meetings
\cite{earliercamop}.
This set the stage for a productive meeting,
running over a period of five days.

\section{Antihydrogen}
The production of antiydrogen in sufficient quantities
for experimental studies holds the potential for new tests of Lorentz
symmetry.
Four groups at the European Organization for Nuclear Research, CERN,
are currently working
on aspects of the production and study of antihydrogen.
They are the
AEGIS (``Antimatter Experiment: Gravity, Interferometry, Spectroscopy"),
ALPHA (``Antihydrogen Laser Physics Apparatus"),
ASACUSA (``Atomic Spectroscopy and Collisions using Slow Antiprotons"),
and ATRAP (``Antihydrogen Trap")
collaborations, all of which were represented at \cpt10.

The AEGIS group is aiming to measure the gravitational acceleration
of antihydrogen,
which would introduce experimental results
into a subject of much theoretical interest
\cite{91Nieto}.
In the initial phase,
the goal is to measure the acceleration at the level of $10^{-2}$
by detecting the free fall of a beam of antihydrogen passed through
a Moir\'e interferometer \cite{2007AEGIS}.
The ALPHA collaboration aims to perform precision tests of CPT symmetry
by comparing the spectra of antihydrogen and hydrogen.
The group reported progress
in the area of
evaporative cooling of antiprotons,
creation of trap conditions in which antihydrogen trapping is realistic,
and the establishment of event selection
criteria \cite{2011Fujiwara}.
Soon after the meeting,
the ALPHA collaboration reported success in trapping antihydrogen atoms
in their apparatus \cite{2010Andresen}.
The ASACUSA collaboration has had success with spectroscopic studies
of antiprotonic helium.
Recent results include an improved measurement of the spin magnetic moment
of the antiproton \cite{asacusa:mu}.
The group has made progress towards
the production of a spin-polarized antihydrogen beam,
which they hope to use for
spectroscopic measurements of the
ground-state hyperfine transitions in antihydrogen atoms
\cite{09Juhasz}.
Such measurements have the potential
to provide clean tests of the CPT symmetry \cite{hbar}.
The ATRAP group has had success
trapping and probing the constituents of antihydrogen.
It was first to demonstrate that antihydrogen could be produced
within a nested Penning and Ioffe-Pritchard trap
configuration \cite{08ATRAP}.

\section{Spectroscopy of atoms}
The exceptional spectroscopic precisions possible with comagnetometers
make them excellent systems for testing Lorentz symmetry.
In these fast-developing experimental systems,
the spin precession rates of several species of atoms
can be compared with exquisite precision.
The information can be used to place limits on
anomalous couplings to Lorentz-breaking background fields.
There are many different background fields,
each coupling to different types of particles.
For example, the $b$-type background fields are relevant
for testing anomalous spin couplings in fermions,
and there are different $b$ coefficients for protons,
neutrons, and electrons.
Recently,
a group at Princeton
placed constraints
at the level of $10^{-33} \gev$ on the equatorial
components of the $b$-type coefficients
in the neutron sector,
a 30-fold improvement on the previous mark
\cite{10Brown}.
These results were achieved with a new apparatus, CPT-II,
that improved on several aspects of the earlier CPT-I
device.
The system, a K-He comagnetomoeter,
is mounted on a rotary platform
which allows for frequent reversals of its orientation.
The main systematic is a coupling
between the Earth's rotation and the gravitational field,
and the team has discussed the interesting possibility
of removing this limitation
by running the experiment near the South pole.

A comagnetometer based on helium and xenon atoms
at the Harvard-Smithsonian Center for Astrophysics
generated a number of the first limits
on neutron coefficients for Lorentz violation
\cite{00Bear,04Cane}.
More recently,
a helium-xenon comagnetometer
at the
Physikalisch-Technische Bundesanstalt (PTB),
the national metrology institute in Berlin,
has been commissioned and is generating competitive limits
\cite{10Gemmel}.

Clock-comparison experiments
are sensitive to
a variety of SME coefficients
\cite{clocksme}
and have led to sharp constraints
on Lorentz violation
using cesium fountain clocks
\cite{clockmaser}
in addition to the devices mentioned above.
Recent work shows that
the reach of these experiments
can be extended by taking into account
various effects,
such as
the binding energies of nucleons,
the velocity of the Earth at different points on its orbit,
and the axial precession of the Earth
\cite{10Altschul}.

A clock-comparison experiment
run at the Laue-Langevin Institute in Grenoble
has the distinction of being the first Lorentz test
based on an experiment with free neutrons.
It compared the spin precession frequencies
of ultra-cold neutrons and mercury atoms,
placing limits on a combination of SME coefficients
\cite{09Altarev}.

\section{Cavity Oscillators and the minimal SME}
Experiments with high-precision optical and microwave oscillators
have shown steady increases in their ability to probe
Lorentz symmetry for a number of years.
These have mainly focused on the
minimal SME
\cite{dcak97,akmm01},
which is made up of
dimension 3 and 4 operators
that break Lorentz symmetry.
In the photon sector,
all the dimension-3 operators,
and 10 of the 19 dimension-4 operators
control birefringence
and are tightly constrained by exploiting the long baselines of
astrophysical observations.
The remaining 9 nonbirefringent operators
in the minimal SME
have been the focus of a number of rapidly-evolving laboratory tests
involving various precision cavity oscillators.
There are at present more than 130
limits on individual SME coefficients
in the minimal photon sector \cite{LVtables},
resulting from work done since 2003.

Presentations by several of the
experimental groups working in this area
were given at \cpt10.
An apparatus at the Humboldt University in Berlin
has two crossed cavity oscillators
and a monolithic optical sapphire resonator.
The design builds on earlier oscillators that have
generated some of the highest resolution tests of Lorentz symmetry
in the photon sector
\cite{09Herrmann}.
Recently, results obtained
at the University of Western Australia,
using an oscillator that operates in the
`whispering gallery' mode,
were used to place
limits on all 9 nonbirefringent coefficients
in the minimal photon sector.
The group has introduced numerous
innovations \cite{09Tobar},
and their latest results include an
improved limit on
one of the isotropic parameters of the SME
\cite{10Hohensee}.
The Australian group
is joining forces with the Humboldt University experimentalists
by bringing their apparatus to Germany.
This will further increase the ability of these
systems to generate refined tests of Lorentz symmetry.
Another German group,
based in D\"usseldorf,
has pushed the precision boundary by
placing limits on SME coefficients
with a cavity experiment investigating the isotropy of the speed of light
\cite{09Eisele}.

Many other physical systems
in the photon sector have played a part
in studies of Lorentz violation,
including, for example,
\v Cerenkov radiation \cite{cerenkov},
nonlinear studies \cite{10alfaro},
the Chern-Simons term \cite{chernSim},
and statics \cite{qbstatics}.

\section{Nonminimal coefficients in the electromagnetic sector}
Several higher-order Lorentz-breaking effects
were presented at the \cpt10 meeting,
based on recent work that has completed
the challenging task of producing a systematic
order-by-order account of all the terms
\cite{NonminAKMM}.
The complete characterization of these coefficients
for Lorentz violation
involves a decomposition in terms of
spin-weighted spherical harmonics.
The formalism is adapted to provide
information about which coefficients
control birefringence, dispersion, and anisotropy
under vacuum or other boundary conditions,
thereby helping to make the distinction between
astrophysical and laboratory tests.
At the \cpt10 meeting,
presentations discussing experimental limits on
nonminimal coefficients included ones made by representatives
for astrophysical tests,
such as ones with the Fermi telescope \cite{2010Vasileiou},
and laboratory tests, as with cavity oscillators \cite{2011parker}.
As of January 2011, there exist
limits on higher-order
photon-sector operators of dimensions
5 through 9,
based on astrophysical birefringence,
cosmic-microwave-background polarization,
and astrophysical dispersion
\cite{LVtables}.

\section{Gravity-sector studies and tests}
The framework for Lorentz violation in curved spacetime
was established several years ago \cite{akgrav04}.
Details of the theory and experimental signals
arising from couplings of the background fields to the pure-gravity sector
have been investigated \cite{qbak06}.
They have led to results on SME coefficients
controlling couplings in the pure-gravity sector,
based on lunar-laser-ranging data \cite{07Battat}
and atom interferometer experiments \cite{09chung,08Muller}.

Recent work on the couplings of the background fields
to the matter sector \cite{akjt}
was presented at \cpt10.
The results indicate a broad range of possibilities
for experimental investigations,
many of which involve atomic, molecular, and optical physics.
Laboratory tests studied
include ones with free-fall and and force-comparison gravimeters,
and others with free-fall and force-comparison
weak-equivalence-principle devices.
Many of the tests are limited by the time of free fall of test masses.
The proposed Sounding-Rocket Principle Of Equivalence Measurement,
SR-POEM,
\cite{JPhillips}
can increase this time up to several hundred seconds.
Weak equivalence principle tests can also be conducted in space,
where test bodies can be compared after almost unlimited periods of free fall.
Several satellite missions have the potential to perform tests,
and include for example the Satellite Test of the Equivalence Principle,
STEP,
which was represented,
together with Gravity Probe-B,
at the \cpt10 meeting \cite{worden}.
Tests of a more exotic type
might include charged-particle interferometry,
comparisons of hydrogen and antihydrogen,
and tests with second- and third-generation particles such as muonium.
Solar-system tests are also of definite interest,
and include lunar and satellite laser ranging,
as well as perihelion precession of Mercury and the Earth.

Another category of tests
would look for gravitational couplings with Lorentz violating background fields
based on the Shapiro time delay,
the gravitational Doppler shift,
and the gravitational redshift.
Recent work at
the University of California, Berkeley,
has focused on the relationship between
different types of gravitational redshift experiments,
and has tested the Einstein equivalence principle
by placing limits
at the level of parts in $10^6$
on coefficients that control spin-independent couplings of
SME fields with the matter sector
\cite{2011equivalencePRL}.

A key result in the fundamental theory of Lorentz violation
is the incompatibility of
explicit Lorentz symmetry breaking with
generic Riemann-Cartan geometries,
a limitation that is evaded by spontaneous Lorentz breaking
\cite{akgrav04}.
Consequently,
a number of studies have been made of theories involving tensor fields
with potentials that trigger spontaneously-broken Lorentz symmetry.
These fields include the
bumblebee $B_\mu$ \cite{akgrav04,2008BluhmEtal},
the symmetric-index cardinal $C^{\mu\nu}$ \cite{2009RPAK},
and an antisymmetric two-tensor $B_{\mu\nu}$ \cite{2010abk}.
These investigations have shown intriguing connections with
Einstein-Maxwell theory, and general relativity,
and have led to a rich body of results relating to
massless and massive propagating modes.

Experiments from the realm of atomic, molecular, and optical physics
have exploited the relationship between Lorentz-breaking fields
and spacetime torsion \cite{torsion}
to place limits on the latter \cite{2008aknrjt}.
Several talks at \cpt10 addressed aspects of torsion,
including its relation to effective field theory \cite{shapiro},
and to rotating Kerr black holes \cite{cambiaso}.
Related work has investigated the classical lagrangians
with Lorentz violation \cite{2010aknr}
that follow from SME-based dispersion relations
\cite{2001Lehnert},
and the associated Riemann-Finsler geometries
\cite{2011finsler,06Bogos,07Girelli}.
Torsion has also been studied as a background \cite{backgrounds},
and in the context of Lorentz violation
and galactic dynamos \cite{2011deandrade}.

\section{Accelerator-based AMO experiments}
A Compton-scattering experiment
at the GRenoble Anneau Acc\'el\'erateur Laser (GRAAL) beamline of the
European Synchrotron Radiation Facility (ESRF)
in Grenoble
has recently led to a sharp test of Lorentz symmetry.
Calculations show that
when a photon is scattered from a free electron
the maximal energy of the backscattered photon,
the Compton-edge energy,
is sensitive to sidereal Lorentz-breaking effects.
Since the sensitivity goes like $\ga^2$,
the ultrarelativistic electrons at the ESRF,
with $\ga=11800$,
are ideally suited to a Lorentz test.
The result \cite{10Bocquet}
is an order of magnitude improvement
on a combination of photon and electron coefficients
in the SME.

Ives-Stilwell experiments,
which measure Doppler shifts of spectral lines,
offer another avenue for tests of Lorentz symmetry
in the matter sector.
Currently,
measurements of optical transitions of lithium ions
in a $\be=0.34$ beam
at the Experimental Storage Ring (ESR)
of the GSI Helmholtz Center for Heavy Ion Research in Darmstadt
are under way.
It was reported at \cpt10
that the sensitivity achieved is on a par with
an earlier experiment at the heavy-ion Test Storage Ring (TSR)
at the Max Planck Institute for Nuclear Physics in Heidelberg,
and that a further order of magnitude improvement is expected.
Bounds on combinations of matter-sector coefficients
for Lorentz violation were presented
\cite{10Saathoff}.

The DA$\Phi$NE collider at Frascati
has the potential to provide experimental results
on SME coefficients for the neutral kaons
\cite{09KLOE}.
Details of the types of signals and analyses
from the Frascati KLOE and KLOE-2
experiments
were presented at \cpt10
\cite{desantis}.

\section{Other sectors}
The SME has had remarkable success in
resolving the anomalies observed
in neutrino experiments in recent years.
The usual three-neutrino Standard Model is not able to account for
recent anomalies \cite{anomalies}
seen in
the Liquid Scintillator Neutrino Detector (LSND) signal,
the Mini Booster Neutrino Experiment (MiniBooNE) low-energy excess,
and the
neutrino-antineutrino differences
in the MiniBooNE and the
Main Injector Neutrino Oscillation Search (MINOS) experiments.
Several talks at \cpt10
addressed neutrino physics,
and included a presentation
of the basic theoretical framework for Lorentz- and CPT-violating
effects in this sector.
The theory demonstrates that differing limits are
relevant for analyses of
short- and long-baseline experiments
\cite{neutrinoSMEtheory}.
Several models have been studied \cite{07BMW}.
Preliminary experimental results from the neutrino sector
presented at the meeting
showed that SME coefficients for Lorentz violation
can be fitted to experimental data from the LSND and MiniBooNE
experiments \cite{katori}.
More recently,
a seven-parameter model,
the  `puma,'
has been built within this SME framework.
It is the first model consistent with all compelling neutrino
data and the LSND, MiniBooNE, and MINOS anomalies \cite{puma}.

Another anomalous result in the context of B-meson oscillations
has been interpreted in terms of the SME,
based on the findings \cite{Banomaly}
of the D$0$ collaboration at FermiLab.
The resulting constraint \cite{2010AKRvK}
at the level of parts in $10^{12}$
is the first sensitivity to CPT violation in the
$B^0_s$ system.

\section{Conclusion}
The energetic and innovative work put forth by the AMO community
in pursuing the possibility of Lorentz violation in nature
continues to lead to technological advances
and the prospect of exciting results.
The {\em Fifth Meeting on CPT and Lorentz Symmetry}
provided a forum for the community to get updated
on developments.
This article emphasizes the AMO physics at the meeting,
although many other topics,
such as
superfields \cite{superfields},
topological defects \cite{defects},
quark condensation \cite{xiong},
linearized gravity \cite{07ferrari},
and coordinate invariance breaking \cite{10coordinvar}
were also discussed.
Experiments continue to allow ever deeper searches
into the details of nature,
while theoretical developments have provided
additional untested regions of coefficient space.
Areas of investigation that are likely to lead to
experimental limits in the coming few years
include coefficients controlling nonminimal couplings in the SME,
and others controlling gravitational couplings.
It is clear that tests of the fundamental symmetries
of nature are profoundly interesting to a broad community of physicists,
and one can hope that new territory will be revealed
as this frontier is further explored.

\end{document}